\documentclass[namedreferences]{SolarPhysics}
\usepackage[optionalrh]{spr-sola-addons} 
\usepackage{graphicx}        
\usepackage{color}           
\usepackage{url}             

\newcommand{\etal}{{\it et al.}}

\newcommand{\be}{\begin{equation}}
\newcommand{\ee}{\end{equation}}
\newcommand{\beq}{\begin{eqnarray}}
\newcommand{\eeq}{\end{eqnarray}}

\newcommand{\aap}{    {\it Astron. Astrophys.}}

\newcommand{\apj}{    {\it Astrophys. J.}}

\newcommand{\solphys}{{\it Solar Phys.}}

\newcommand{\va}{v_{\rm A}}
\newcommand{\cs}{c_{\mathrm{s}}}
\newcommand{\ct}{c_{\mathrm{T}}}

\newcommand{\vap}{v_{\mathrm{Ap}}}
\newcommand{\csp}{c_{\mathrm{sp}}}
\newcommand{\ctp}{c_{\mathrm{Tp}}}

\newcommand{\vapc}{v_{\mathrm{Atr}}}
\newcommand{\cspc}{c_{\mathrm{str}}}
\newcommand{\ctpc}{c_{\mathrm{Ttr}}}

\newcommand{\vac}{v_{\mathrm{Ac}}}
\newcommand{\csco}{c_{\mathrm{sc}}}
\newcommand{\ctc}{c_{\mathrm{Tc}}}

\newcommand{\kzp}{k_{z{\rm p}}}
\newcommand{\kzc}{k_{z{\rm c}}}
\newcommand{\kzpc}{k_{z{\rm tr}}}

\begin{document}
\begin{article}
\begin{opening}

\title{Oscillatory Modes of a Prominence-PCTR-Corona Slab Model}

\author{R.~\surname{Soler}\sep 
        R.~\surname{Oliver}\sep 
        J.L.~\surname{Ballester}}

\runningauthor{R. Soler, R. Oliver, and J.L. Ballester}
\runningtitle{Oscillatory Modes of a Prominence-PCTR-Corona Slab Model}

\institute{R.~\surname{Soler}\sep 
        R.~\surname{Oliver}\sep 
        J.L.~\surname{Ballester} \\ Departament de F\'isica, Universitat de les Illes Balears, E-07122, Palma de Mallorca, Spain \\
                  email: \url{roberto.soler@uib.es} \\ email: \url{ramon.oliver@uib.es} \\ email: \url{joseluis.ballester@uib.es}
            }

\date{Received: 19 January 2007 / Accepted: 16 October 2007}


\begin{abstract}

Oscillations of magnetic structures in
the solar corona have often been interpreted in terms of magnetohydrodynamic waves. We study the
adiabatic magnetoacoustic modes of 
a prominence plasma slab with a uniform longitudinal magnetic
field, surrounded by a prominence-corona transition region (PCTR) and a coronal medium. Considering
linear small-amplitude oscillations, the dispersion relation for the
magnetoacoustic slow and fast modes is deduced assuming evanescent-like
perturbations in the coronal medium. In the system without PCTR, a classification of the oscillatory modes
according to the polarisation of their eigenfunctions is made in order to
distinguish modes with fast-like or slow-like properties. Internal and external slow modes are governed by the prominence and coronal properties respectively, and fast modes are mostly dominated by prominence conditions for the observed wavelengths. In addition, the inclusion of an isothermal PCTR does not substantially influence the mode frequencies, but new solutions (PCTR slow modes) are present.

\end{abstract}

\end{opening}


\section{Introduction}
\label{Introduction} 

Despite recent observational and theoretical advances (see the reviews by
\opencite{martin}; \opencite{patso}), the intimate physical nature of solar
prominences is still enigmatic. It is clear that prominences can only exist at
locations where the coronal magnetic field is ``peculiar'' enough to allow for
the equilibrium of the dense prominence plasma against the force of gravity.
Nevertheless, there remain many unknowns in the birth, evolution, and death of
these objects.

Prominence seismology attempts to contribute to the
research of prominence properties. Observations of waves and oscillations in
prominences have become increasingly complex in the last decade and a rich field
has emerged. Oscillatory periods ranging from less than a minute
\cite{balthasar} to 12 hours \cite{foullon} have been detected. It has also
been found that perturbations are typically attenuated with a damping time of a
few periods \cite{terradas02} and in some cases the phase speed and wavelength
of propagating perturbations have also been characterised ({\it e.g.} \opencite{terradas02}). Finally, although there is a
considerable gap between observations and theory, some attempts of doing
prominence seismology have been undertaken and reasonable values of the
prominence physical parameters have been obtained \cite{pouget}. The
observational background has been reviewed at length in various publications
\cite{oliverballester02,wiehr,engvold,banerjee} to which the
reader is referred for more information.

The theoretical magnetohydrodynamic models developed so far start from very crude representations of
the prominence and usually focus on a few ``ingredients'' whose relevance wants
to be tested. For example, \inlinecite{joarderroberts93} studied the adiabatic modes of
oscillation of a plasma slab threaded by a skewed magnetic field. In
\inlinecite{oliverballester96} the magnetic geometry was simplified by taking a transverse
magnetic field to the prominence slab, but a prominence-corona transition region
(PCTR) was included; this is the only theoretical study of prominence oscillations in which the influence of the PCTR has been taken into account. The temporal and spatial  damping of perturbations has been
investigated in some other works ({\it e.g.} \opencite{terradas05}). \inlinecite{oliverballester02} and \inlinecite{banerjee} kave reviewed this subject.

The purpose of this work is to address the importance of the PCTR on the oscillations of a prominence slab threaded by a straight, uniform magnetic field parallel to the slab axis (and not perpendicular to it as in \opencite{oliverballester96}). The PCTR is a narrow layer through which the physical parameters (temperature, density, ionisation degree, {\it etc.}) change abruptly from prominence to coronal values \cite{cirigliano}. The availability of telescopes for the observation of spectral lines sensitive to PCTR temperatures has allowed the detection of waves in this region \cite{blanco, bocchialini, pouget}. The first of these works reports the propagation of a disturbance travelling at $\approx 170$~km~s$^{-1}$ and with a period in the band 3.2\,--\,6.4~minutes. \inlinecite{pouget} reported periodicities detected in three solar filaments
in the He\,{\sc i} line at 5843.3~\AA~(corresponding to 20\,000~K) and identified some of
the periods with the six fundamental oscillatory modes according to the model
proposed by \inlinecite{joarderroberts93}. Incorporating the PCTR in our model will allow us to test whether it has an observable effect on the oscillations of a prominence.

When the PCTR is removed, our equilibrium configuration is identical to that of \inlinecite{ER82}, hereafter ER82. ER82 studied a wide range of values for the physical
conditions of the slab and the outer medium, plotting dispersion diagrams for
each situation, and also discussed the dispersion
relation in the long- and short-wavelength limits. \inlinecite{JR92}, hereafter JR92, extended the work of ER82 by considering the
possibility of motions and propagation in the $y$-direction and
made a more refined classification of the oscillation modes according to their
phase speed. Our purpose is to revise the work done by JR92 in the case
 of no motions and no propagation in the $y$-direction (i.e. the case analysed by ER82), taking into account realistic physical conditions similar to
those in a prominence and its surrounding PCTR and corona, and to study the
polarisation of motions and the other perturbations. Additionally, we
explore the parameter space in an attempt to understand how the variation of the
temperature, density and magnetic field affects the oscillatory period.

This paper is organised as follows: Section~2 contains a description of the
equilibrium model and the linear, adiabatic wave equations, besides the derivation of the 
dispersion relation of magnetoacoustic modes. The
oscillatory modes are studied first for a system without
PCTR (as a revision of ER82) in Section~3. Next, an isothermal PCTR is included in the equilibrium and its effects are investigated in Section~4. Finally, our
conclusions are given in Section~5.


\section{Equilibrium, Basic Equations, and Dispersion Relation \label{sec:equations}} 

\begin{figure}[!t]     
\centering
\includegraphics[width=0.85\columnwidth]{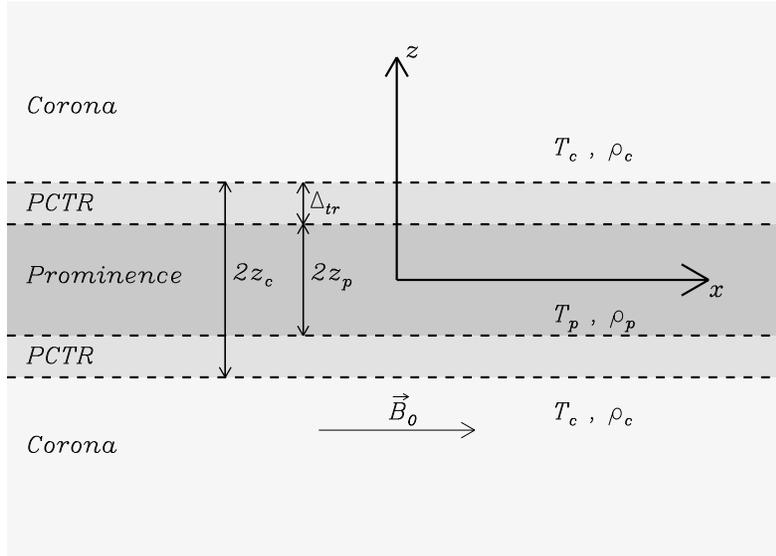}
\caption{Sketch of the equilibrium.}
\label{fig:equilibrium} 
\end{figure}

Our equilibrium configuration (Figure~\ref{fig:equilibrium}) is made of an homogeneous plasma layer with prominence conditions embedded in an unbounded corona with a prominence-corona transition region between them. The PCTR is represented as an isothermal and homogeneous layer. The physical conditions (two different temperatures for the PCTR have been considered) together with the characteristic velocities of each medium are displayed in Table~\ref{tab:cond}. PCTR and coronal densities are computed by fixing their respective temperatures and imposing pressure continuity across the interfaces. The magnetic field is orientated along the $x$-direction, having a value of $5~\mathrm{G}$ in all layers. The three media are unlimited in the $x$- and $y$-directions. The half-width of the prominence slab is $z_{\rm p} = 3000\, \mathrm{km}$ and the PCTR width is $\Delta_{\rm tr} = 900\, \mathrm{km}$.




In our study of linear, adiabatic, magnetohydrodynamic waves we follow the derivation of ER82 (see also \opencite{roberts81}) and use their Equation~(5), which governs the propagation of magnetoacoustic fast and slow modes. The Cartesian axes have a different orientation in ER82 and here, so ER82's Equation~(5) is now written

\begin{equation}
\frac{{\rm d}^2 v_z}{{\rm d} z^2} + k_z^2 v_z = 0, \label{21}
\end{equation}
with
\begin{equation}
k_z^2 = \frac{\left( \omega^2 - k_x^2 \va^2 \right) \left( \omega^2 - k_x^2 \cs^2 \right)}
{\left( \va^2 + \cs^2 \right) \left( \omega^2 - k_x^2 \ct^2 \right)}, \label{22}
\end{equation}
where $v_z$ is the normal velocity component, $\omega$ is the frequency, $k_x$ and $k_z$ are the components of the wavenumber, $\cs^2 = \frac{\gamma p_0}{\rho_0}$ is the adiabatic sound speed squared, $\va^2 = \frac{B_0^2}{\mu \rho_0}$ is the Alfv\'en speed squared and $\ct^2$ is the cusp (or tube) speed squared,
\begin{equation}
\ct^2 = \frac{\va^2 \cs^2}{\va^2 + \cs^2}. \label{23}
\end{equation}
The parameters $B_0$, $p_0$, and $\rho_0$ represent the equilibrium magnetic field strength, pressure, and density of any region. In all the following formulae in this work, subscripts ``p'', ``tr'', or ``c'' denote quantities computed using prominence, PCTR, or coronal values, respectively. 

The expressions for the perturbed quantities as a function of $v_z$ and its derivative are
\begin{eqnarray}
v_x &=& \frac{- i k_x \cs^2}{\omega^2 - k_x^2 \cs^2} \frac{{\rm d} v_z}{{\rm d} z}, \\
\rho_1 &=& \frac{i \omega \rho_0}{\omega^2 - k_x^2 \cs^2}\frac{{\rm d} v_z}{{\rm d} z}, \\
p_1 &=& \frac{i \omega \rho_0 \cs^2}{\omega^2 - k_x^2 \cs^2}\frac{{\rm d} v_z}{{\rm d} z}, \\
B_{1x} &=& \frac{i B_0}{\omega} \frac{{\rm d} v_z}{{\rm d} z}, \\
B_{1z} &=& \frac{B_0 k_x}{\omega}  v_z,
\end{eqnarray}
with $v_x$ the velocity perturbation along the magnetic field and $\rho_1$, $p_1$, $B_{1x}$, and $B_{1z}$ the perturbations to the density, pressure and the $x$- and $z$-components of the magnetic field. Next, we write expressions for the perturbations to the magnetic pressure ($p_{1 \rm m}$) and the total pressure ($p_{1 \rm T}$),
\begin{eqnarray}
p_{1 \rm m} &=& \frac{B_0}{\mu} B_{1x} = \frac{i \rho_0 \va^2}{\omega} \frac{{\rm d} v_z}{{\rm d} z}, \\
p_{1 \rm T} &=& p_1 + p_{1 \rm m} =  \frac{i \rho_0 \left( \omega^2 - k_x^2 \va^2 \right)}{\omega k_z^2} \frac{{\rm d} v_z}{{\rm d} z}.
\end{eqnarray}


\begin{table}[!t]
\begin{tabular}{lrrrrrr}
\hline & $T_0\, (\mathrm{K})$ & $\rho_0\, (\mathrm{kg}\, \mathrm{m}^{-3})$ & $\tilde{\mu}$ & $\ct$ & $\cs$ & $\va$ \\
\hline Prom. & 8\,000 & $5.00 \times 10^{-11}$ & 0.8 & 11.56 & 11.76 & 63.08 \\
PCTR (1) & 50\,000 & $5.00 \times 10^{-12}$ & 0.5 & 36.56 & 37.19 & 199.47 \\
PCTR (2) & 500\,000 & $5.00 \times 10^{-13}$ & 0.5 & 115.62 & 117.62 & 630.78 \\
Corona & 1\,000\,000 & $2.50 \times 10^{-13}$ & 0.5  & 163.51 & 166.33 & 892.06 \\
\hline
\end{tabular}
\caption{Temperature, density, and mean atomic weight together with the cusp, sound, and Alfv\'en speeds for all the considered media. Velocities are expressed in $\mathrm{km}\, \mathrm{s}^{-1}$.}
\label{tab:cond}
\end{table}

In order to obtain an analytical dispersion relation for the linear adiabatic magnetoacoustic waves, we must impose a solution that satisfies Equations~(\ref{21}) and (\ref{22}) for our equilibrium. Since we are interested in wave modes which are evanescent in the corona we impose the restriction $\kzc^2 < 0$. Regarding the prominence, solutions with $\kzp^2 > 0$ correspond to body waves, whereas solutions with $\kzp^2 < 0$ are surface-like modes. On the other hand, in the transition region the case $\kzpc^2 > 0$ corresponds to a propagating wave, whereas the case $\kzpc^2 < 0$ corresponds to an evanescent, tunnelling wave in the PCTR. All of these cases are allowed and studied in this investigation. The analytical function proposed for $v_z$ as solution of Equation~(\ref{21}) is then a piece-wise function,
\begin{equation}
  v_z(z) = \left\{
  \begin{array}{lcrcccl}
  A_1 \exp \left[ \kzc \left( z + z_{\rm c} \right) \right], & \mathrm{if} & & & z & \leq & -z_{\rm c}, \\
  A_2 \cos \left[ \kzpc \left( z + z_{\rm p} \right) \right] + A_3 \sin \left[ \kzpc \left( z + z_{\rm p} \right) \right], & \mathrm{if} & -z_{\rm c} & < & z & < & -z_{\rm p}, \\
  A_4 \cos \left( \kzp\, z \right) + A_5 \sin \left( \kzp\, z \right), & \mathrm{if} & -z_{\rm p} & \leq & z & \leq & z_{\rm p}, \\
  A_6 \cos \left[ \kzpc \left( z - z_{\rm p} \right) \right] + A_7 \sin \left[ \kzpc \left( z - z_{\rm p} \right) \right], & \mathrm{if} & z_{\rm p} & < & z & < & z_{\rm c}, \\
  A_8 \exp \left[- \kzc \left( z - z_{\rm c} \right) \right], & \mathrm{if} & z_{\rm c} & \leq & z. &  & 
  \end{array} \right. \label{24}
\end{equation}
This solution satisfies the evanescent assumption, $v_z \to 0$ when $z \to \pm \infty$. Imposing continuity of $v_z$ and the total pressure perturbation across the interfaces, we obtain eight homogeneous linear algebraic equations for the constants $A_1-A_8$. The non-trivial ({\it i.e.} non-zero) solution of this system gives us the dispersion relation
\begin{equation}
\frac{\rho_{\rm tr}}{ \rho_{\rm p}} \left( k_{x}^{2} \vapc^2 - \omega^{2}\right) \kzp 
 \left\{ \begin{array}{l} \cot \\ \tan \end{array} \right\}  \left( \kzp z_{\rm p} \right) \mathcal{R}_1 \pm \left( k_{x}^{2} \vap^{2} - \omega^{2}\right) \kzpc \mathcal{R}_2=0, \label{disperpctr}
\end{equation}
with
\begin{eqnarray}
\mathcal{R}_1 &=& \rho_{\rm c}  \kzpc \left( k_{x}^{2} \vac^2 - \omega^{2}\right) \cos \left[ \kzpc \left( z_{\rm c} - z_{\rm p} \right) \right] + \nonumber \\ &&  \rho_{\rm tr}  \kzc \left( k_{x}^{2} \vapc^2 - \omega^{2}\right) \sin \left[ \kzpc \left( z_{\rm c} - z_{\rm p} \right) \right], \\
\mathcal{R}_2 &=& \rho_{\rm tr}  \kzc \left( k_{x}^{2} \vapc^2 - \omega^{2}\right) \cos \left[ \kzpc \left( z_{\rm c} - z_{\rm p} \right) \right] - \nonumber \\ &&  \rho_{\rm c}  \kzpc \left( k_{x}^{2} \vac^2 - \omega^{2}\right) \sin \left[ \kzpc \left( z_{\rm c} - z_{\rm p} \right) \right],
\end{eqnarray}
where $\cot$/$\tan$ terms and $\pm$ signs in Equation~(\ref{disperpctr}) are related with the symmetry of the perturbations: the $\cot$ term and $+$ sign correspond to kink modes ($A_5=0$), while the $\tan$ term and $-$ sign correspond to sausage modes ($A_4=0$). When the PCTR is removed from the equilibrium ($z_{\rm c} = z_{\rm p}$), the dispersion relation (\ref{disperpctr}) reduces to
\begin{equation}
\frac{\rho_{\rm c}}{ \rho_{\rm p}} \left( k_{x}^{2} \vac^2 - \omega^{2}\right) \kzp 
\left\{ \begin{array}{l} \cot \\ \tan \end{array} \right\}  
\left( \kzp z_{\rm p} \right) \pm \left( k_{x}^{2} \vap^{2} - \omega^{2}\right) \kzc =0. \label{dispernopctr}
\end{equation}
This dispersion relation is equivalent to Equation~(11) in ER82.



\section{Results: Equilibrium without PCTR}
\label{res:nopctr}

First, we consider an equilibrium without PCTR and take $k_y = 0$, which is equivalent to that of ER82. The study of this simplified system allows us to understand the basic solutions of the dispersion relation and their behaviour, which will be useful in order to compare with the results obtained when a PCTR is included in the system. Since ER82 and JR92 investigated in detail this equilibrium, we concentrate our attention in the study of the oscillatory modes according to the polarisation of their eigenfunctions and to their response when the equilibrium physical parameters (temperature, density, magnetic field, {\it etc.}) are changed. 

\subsection{Phase Speed Diagram}

\begin{figure}[!h]
\centering
\includegraphics[width=0.85\columnwidth]{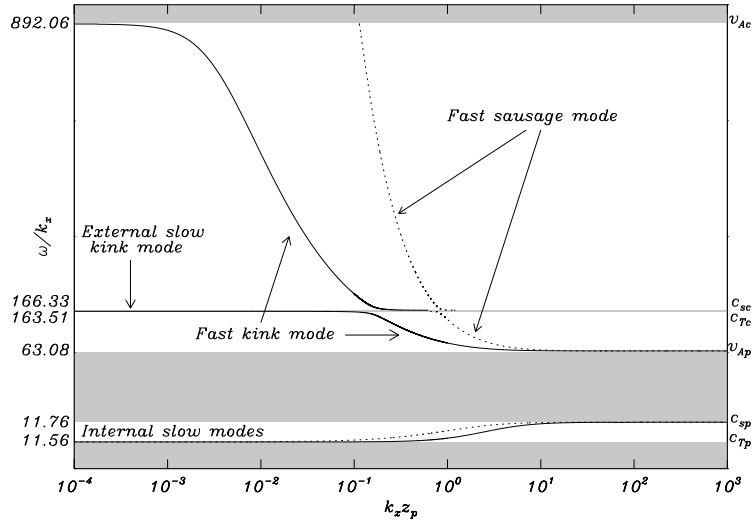}
\caption{Phase speed (in km~s$^{-1}$) versus the dimensionless longitudinal wavenumber of the normal modes of a prominence-corona system without transition region. Solid lines denote kink modes and dotted lines denote sausage modes. Only the fundamental modes are plotted. The shaded zones are forbidden regions where evanescent-like solutions in the corona do not exist. Note that the vertical axis is not drawn to scale.}
\label{fig:adiabatic}
\end{figure}

In Figure~\ref{fig:adiabatic} the magnetoacoustic wave dispersion diagram is
displayed. As noted in ER82, the present ordering of Alfv\'en and sound
speeds implies that externally evanescent solutions only exist in three phase
speed ranges (or windows): $\ctp < \omega/k_x < \csp$, $\vap < \omega/k_x < \ctc$
and $\csco < \omega/k_x < \vac$. Each window contains an infinite number of
harmonics, although we restrict our study to the fundamental kink and sausage
modes and so only two curves are represented in each window of
Figure~\ref{fig:adiabatic}. In the case $k_y = 0$, JR92 denoted that the modes in
the three phase speed ranges as slow, fast externally slow and fast externally fast (or fast
ES and fast EF for short), respectively. It is worth mentioning that for $k_y = 0$ and for the present ordering of Alfv\'en and sound
speeds this equilibrium does not support surface-like modes ($\kzp^2 < 0$) since all solutions are of the form of body waves ($\kzp^2 > 0$). Notice that both ER82 and JR92 missed the horizontal ({\it i.e.} non-dispersive) solutions in the top two windows of the phase speed diagram (see their Figure~7 and Figure~2(a), respectively).  The study of the nature of
the solutions carried out in the next two subsections suggests that the classification of JR92 for $k_y = 0$ might not fully reflect the physical nature of the modes. A more
representative description of the magnetoacoustic modes is as follows:

\begin{itemize}

\item Solutions in the bottom window are internal slow modes, that exist for all
values of the longitudinal wavenumber ($k_x$) and whose features are mainly
determined by the prominence physical properties. The phase speed of the internal slow modes is enclosed in a very narrow region between $\ctp$ and $\csp$, so a good approximation for the frequency is $\omega \approx \csp k_x$.

\item The dispersive solutions in the upper two windows are the kink and sausage fast
modes. The phase speed of fast modes varies between $\vap$ and $\vac$, but they
are separated in two different windows by the ``forbidden region'' ($\ctc < \omega/k_x < \csco$),
which causes the presence of lower and upper cut-off frequencies. In this forbidden region, evanescent waves are not possible since solutions become leaky ($\kzc^2 > 0$), which means that they freely propagate in the coronal medium and carry energy away from the prominence slab. The ``forbidden region'' causes the appearance of an horizontal branch with $\omega/k_x\approx\ctc$, which corresponds to a solution whose properties are slow-like and fixed by the coronal medium, as we explain later in Section~\ref{sec:param}. For this reason, we call this branch of the solution the external slow kink mode, although one must bear in mind that this mode is not propagating in the corona and the word ``external'' only points out that its behaviour is dominated by the properties of the outer medium. In Figure~\ref{fig:coupling} we show an enlargement of the zone in the dispersion diagram close the forbidden region, where the fundamental fast kink mode separates into two branches and the external slow kink mode arises.

\begin{figure}[!h]
\centering
\includegraphics[width=0.85\columnwidth]{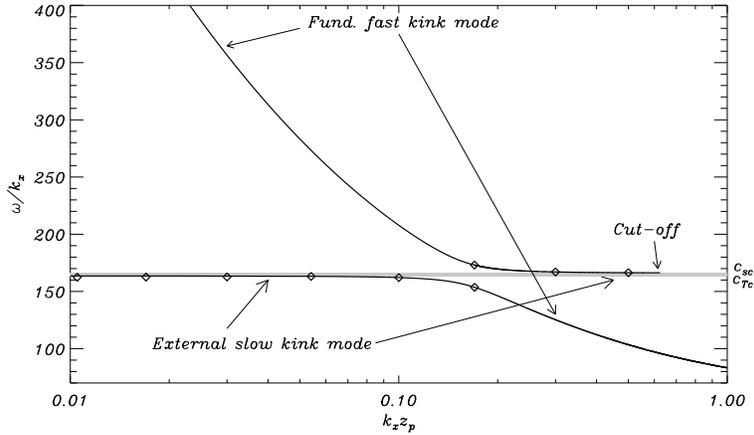}
\caption{Enlargement of Figure~\ref{fig:adiabatic} to show the distinction between the fundamental fast kink mode (solid line) and the external slow kink mode (solid line and symbols). Note the cut-off of the external slow kink mode. The phase speed is given in km~s$^{-1}$.}
\label{fig:coupling}
\end{figure}

\end{itemize}

  In addition, Figures.~\ref{fig:adiabatic} and \ref{fig:coupling} show the presence of a cut-off frequency in the upper branch of the external slow kink mode: this solution is only evanescent for $k_x z_p \lesssim 0.6$ and becomes leaky ($\kzc^2 > 0$) for $k_x z_p \gtrsim 0.6$. Therefore, it is not drawn in Figures.~\ref{fig:adiabatic} and \ref{fig:coupling} for $k_x z_p \gtrsim 0.6$. The cut-off frequency is present in the upper branch of the external slow mode and not in the lower branch due to the fact that its phase speed tends to cluster towards the external tube speed, $\ctc$. The lower branch approaches $\ctc$ as an evanescent-like solution since the leaky region is not crossed. On the other hand, the upper branch of the external slow solution becomes leaky because the sign of $\kzc^2$ changes from negative to positive when its phase speed reaches the external sound speed (see Equation~(\ref{22})) in its way to the clustering towards $\ctc$.

\subsection{Properties of the Oscillatory Modes}
\label{sec:eigen}

\begin{figure}[!h]
\centering
\includegraphics[width=\columnwidth]{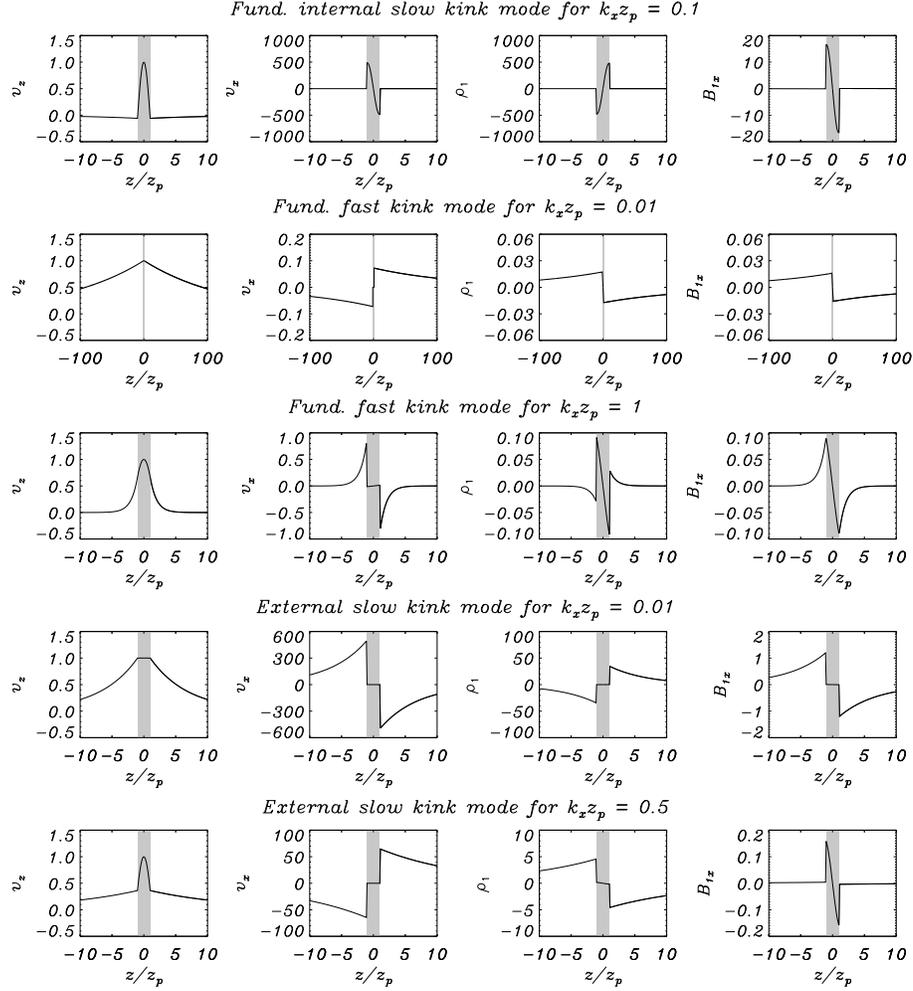}
\caption{Perturbations $v_z$, $v_x$, $\rho_1$, and $B_{1x}$ (in arbitrary units) versus the dimensionless distance to the prominence axis corresponding to the fundamental kink modes for some selected values of $k_x z_{\rm p}$. The shaded region corresponds to the prominence slab. The plots in the second row (fast mode for $k_x z_{\rm p} = 0.01$) have a different scale for the horizontal axis to emphasize the poor confinement of this solution.} 
\label{fig:autofunctions}
\end{figure}

In order to study in depth the fundamental oscillatory modes and their magnetoacoustic properties, we plot the eigenfunctions for some selected values of the wavenumber. In particular, we pay attention to the perturbed velocities ($v_z$ and $v_x$), the perturbed density ($\rho_1$), and the $x$-component of the perturbed magnetic field ($B_{1x}$) (see Figure~\ref{fig:autofunctions}). The behaviour of non-leaky kink and sausage solutions is similar, so we concentrate on the former.

Beginning with internal slow modes, they do not have cut-off frequencies and so they exist for all values of $k_x z_{\rm p}$.  The eigenfunctions corresponding to the fundamental slow kink mode are plotted only for $k_x z_{\rm p} = 0.1$ in Figure~\ref{fig:autofunctions} (top row) since they are similar for other values of $k_x z_{\rm p}$. We can observe that perturbations are essentially confined to the slab, and the dominant component of the perturbed velocity is the one parallel to the magnetic field ($v_x$). On the other hand, $\rho_1$ and $B_{1x}$ are also large in the slab, as for slow waves. Since perturbations are much larger in the prominence than in the corona, we expect that internal slow modes are dominated by prominence conditions whereas the corona has little or no influence.

Regarding fast modes, they are dispersive and their phase speed varies between $\vap$ and $\vac$. All of these modes, except for the fundamental kink, have an upper cut-off frequency ($\omega _{\rm cut}$). For $\omega > \omega _{\rm cut}$, modes become leaky. Moreover, additional cut-off frequencies occur due to the presence of the forbidden region \mbox{$\ctc < \omega/k_x < \csco$}. The eigenfunctions of the fundamental fast kink mode for \mbox{$k_x z_{\rm p}  = 0.01$}  and $k_x z_{\rm p}  = 1$ can be compared in Figure~\ref{fig:autofunctions} (second and third rows). The confinement of perturbations to the prominence slab is poorer than for the slow modes and decreases with decreasing $k_x z_{\rm p}$. The perturbations for small $k_x z_{\rm p}$ have long tails which penetrate far into the corona, so we expect that these solutions are highly influenced by coronal conditions. This is in agreement with \inlinecite{diaz01}, who studied fast oscillations in a Cartesian line-tied slab and found that the amplitude of transverse oscillations can still be large in the corona at a distance of many slab widths. Such as corresponds to a fast mode, the dominant component of the velocity is $v_z$ and the density and longitudinal magnetic field perturbations are much smaller than in the case of slow modes. For $k_x z_{\rm p}  = 1$, both components of the velocity have a similar amplitude in the corona because the fast mode is externally slow. However, $v_z$ is dominant in the corona for $k_x z_{\rm p}  = 0.01$ because the fast mode is now externally fast.

Finally, the perturbations for the external slow mode are plotted for $k_x z_{\rm p} = 0.01$ and $k_x z_{\rm p} = 0.5$. This solution presents a strong, slow-like polarisation outside the slab, with the amplitude of $v_x$ much larger than that of $v_z$, and produces significant density and longitudinal magnetic field perturbations compared to those of the fast mode. Nevertheless, the polarisation of oscillations in the prominence slab is fast-like, {\it i.e.} $v_z$ is larger, but since the amplitude of perturbations is much larger in the corona than in the slab, one expects that coronal properties govern the behaviour of this solution, its slow-like character being therefore dominant. This is verified in the next Section.

\subsection{Dependence on the Equilibrium Physical Conditions}
\label{sec:param}

\begin{figure}[!h]
\centering
\includegraphics[width=\columnwidth]{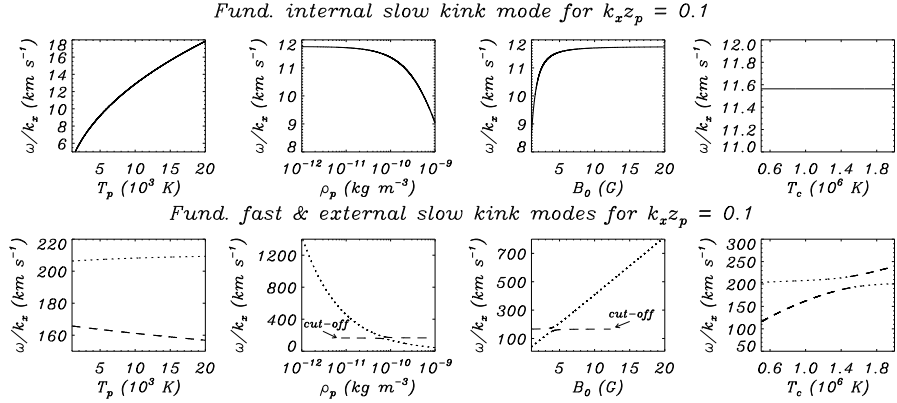}
\caption{Phase speed of the fundamental slow kink mode with $k_x z_{\rm p} = 0.1$ (solid line in the top row) and the fundamental fast and external slow kink modes with $k_x z_{\rm p} = 0.1$ (dotted and dashed lines in the bottom row, respectively). From left to right the four columns display the variation of the phase speed against the prominence temperature, the prominence density, the magnetic field and the coronal temperature.} 
\label{fig:parameters}
\end{figure}

Next, we investigate how the phase speed of the fundamental modes is affected by changing the equilibrium physical parameters. For simplicity, we limit this analysis to kink modes (Figure~\ref{fig:parameters}).

In the case of the internal slow mode for $k_x z_{\rm p} = 0.1$, the phase speed only depends on the prominence physical conditions whereas the value of the coronal temperature does not affect it. The dependence of the phase speed on the temperature can be understood from the above mentioned approximation for the internal slow mode frequency ($\omega  \approx k_x \csp$), leading to $v_{\rm ph} \sim \csp$, so $v_{\rm ph} \sim T_{\rm p}^{1/2}$. Finally, we see that the phase speed of slow modes decreases for weak field and large density. This fact can be understood taking into account that the phase speed diagram presented in Figure~\ref{fig:adiabatic} is modified for weak magnetic field and large density. In this situation, $\vap$ becomes smaller than $\csp$ and then the windows of existence of non-leaky fast and internal slow modes overlap, and both modes couple. This causes the behaviour of the internal slow mode phase speed for weak field and large density.

Regarding the fast and the external slow modes, their phase speeds for $k_x z_{\rm p} = 0.1$ are plotted together in Figure~\ref{fig:parameters} (bottom panels). The fast mode is almost independent of prominence and coronal temperatures, and the weak dependence that is shown is due to the closeness of the coupling with the external slow mode. The values of the prominence density and magnetic field affect the fast mode in an important way. The phase speed increases with the magnetic field in a linear way, whereas it decreases when large values of the prominence density are considered. Both dependencies are consistent with the rough approximation $v_{\rm ph} \sim \vap$, so $v_{\rm ph} \sim B_0$ and $v_{\rm ph} \sim \rho_{\rm p}^{-1/2}$.

Finally, the external slow mode only depends on the temperatures, but it is unaffected by the value of the prominence density and magnetic field. The phase speed slightly decreases when the prominence temperature grows, whereas it grows dramatically with the coronal temperature according to the approximate relation $v_{\rm ph} \sim T_{\rm c}^{1/2}$, so one can conclude that the corona is the dominant medium regarding the properties of this mode, as we suggested in Section~\ref{sec:eigen}. In fact, the external slow mode behaves as the slow mode in an unbounded medium with coronal properties, although we must bear in mind that it is not a propagating but an evanescent wave in the external medium. In addition, this solution does not exist for small density and strong field because the external tube speed, $\ctc$, is then less than $\vap$  under these conditions.


\section{Results: Equilibrium with an Isothermal PCTR}
\label{res:pctr}

\subsection{Modification of the Phase Speed Diagram}

Now, we include a PCTR in the equilibrium configuration. Our aim is to study the effects arising from the presence of the PCTR on the basic oscillatory modes of the system without PCTR, which have been investigated and classified in the last section. Since we assume an isothermal transition between the prominence and coronal media, the PCTR temperature ($T_{\rm tr}$) is a relevant parameter. One must bear in mind that to consider an isothermal PCTR is, obviously, a rough approximation but it allows us to obtain an analytical expression for the dispersion relation.

\begin{figure}[!p]
\centering
\includegraphics[width=0.85\columnwidth]{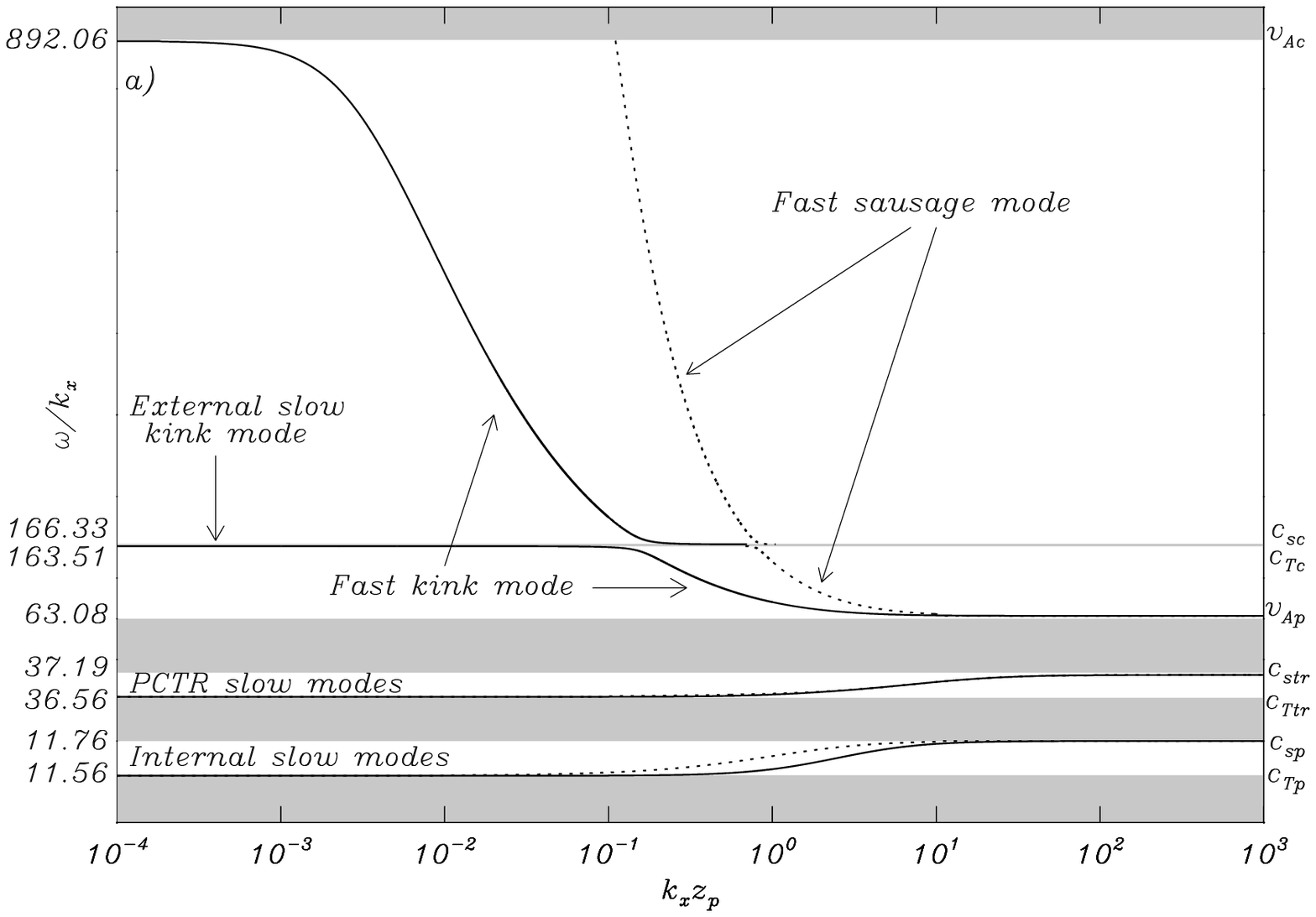}\\
\includegraphics[width=0.85\columnwidth]{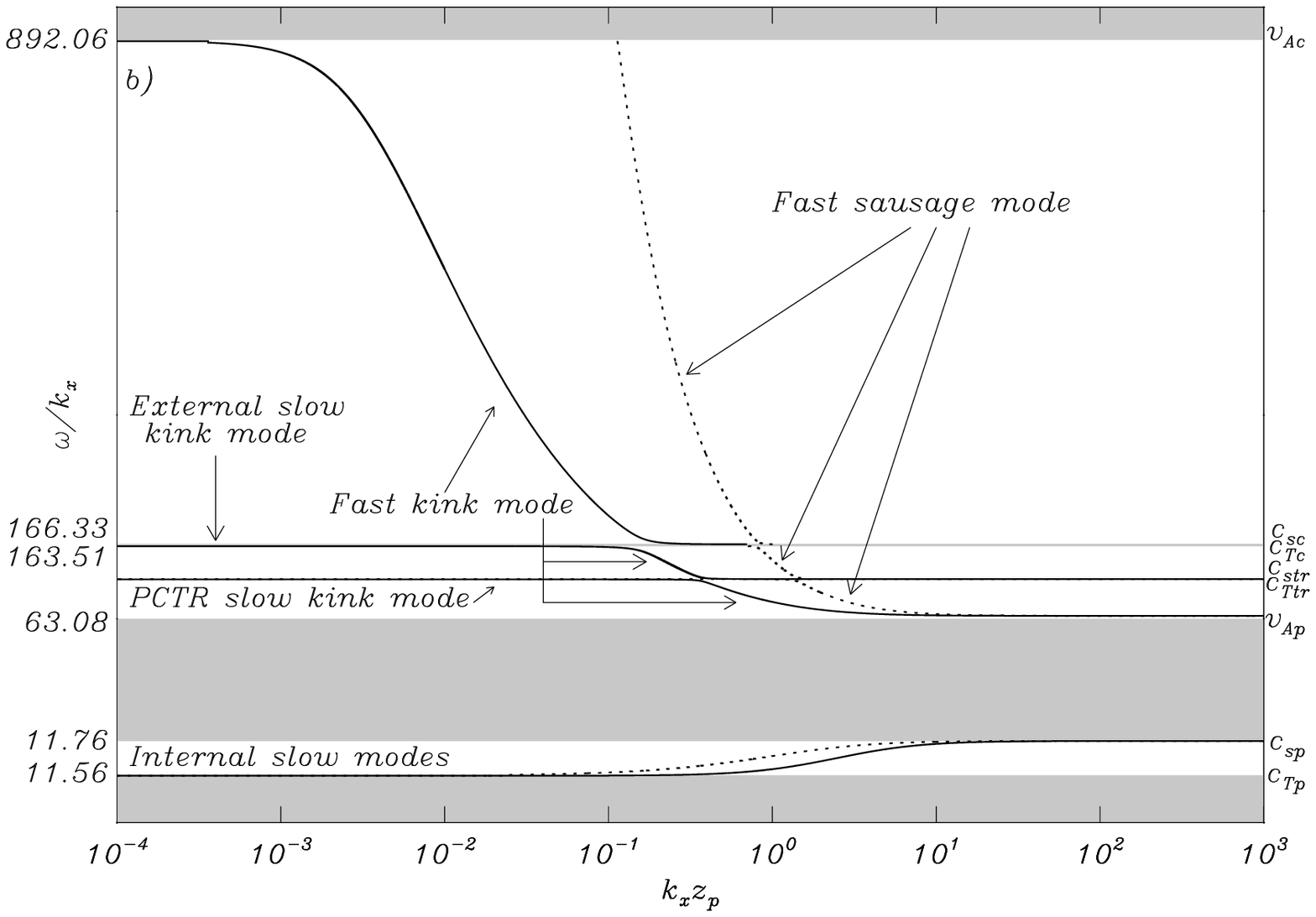}
\caption{Phase speed (in km~s$^{-1}$) versus the dimensionless longitudinal wavenumber of the normal modes of a prominence-corona system with transition region. Solid lines denote kink modes and dotted lines denote sausage modes. Only the fundamental modes are plotted. The shaded zones are forbidden regions where evanescent-like solutions of the dispersion relation do not exist. Computations performed for a PCTR temperature of $a)$ 50\,000~K and $b)$ 500\,000~K.} 
\label{fig:pctrdiag}
\end{figure}

When a PCTR is present in the system, solutions are propagating in the PCTR ($\kzpc^2 > 0$) if their phase speed lies in the region \mbox{$\ctpc < \omega/k_x < \cspc$}. Outside this region, solutions are tunnelling in the PCTR ($\kzpc^2 < 0$). Depending on the value of the PCTR temperature, the character of the solutions can be different, such as is summarised next.
\begin{itemize}
\item If $\cspc < \vap$ ($T_{\rm tr} <$~144\,000~K for our equilibrium parameters), a new set of solutions of the dispersion relation (Equation~(\ref{disperpctr})) is found in the region \mbox{$\ctpc < \omega/k_x < \cspc$}, which falls into the zone $\csp < \omega/k_x < \vap$, where the solutions are surface-like in the prominence \mbox{($\kzp^2 < 0$)}. This is the only case in which surface waves are possible in the prominence for $k_y = 0$.
\item On the contrary, for $\ctpc > \vap$ ($T_{\rm tr} >$~149\,000~K) the region $\ctpc < \omega/k_x < \cspc$ is embedded into the middle propagating window in the dispersion diagram, and so solutions with \mbox{$\ctpc < \omega/k_x < \cspc$} are now body-like ($\kzp^2 > 0$) in the prominence. 
\end{itemize}
For simplicity, we define $T_{\rm tr}^*$ as the PCTR temperature which separates one situation from the other. It is worth mentioning that an intermediate situation exists when $\ctpc < \vap < \cspc$ and although it could be of academic interest, it corresponds to a very peculiar case and is not examined here. Since taking a uniform $T_{\rm tr}$ in the equilibrium is an idealisation, both $T_{\rm tr} < T_{\rm tr}^*$ and $T_{\rm tr} > T_{\rm tr}^*$ are equally (un)realistic and so both must be considered. Nevertheless, the presence of a PCTR with $T_{\rm tr} < T_{\rm tr}^*$ does not modify in an important way the phase speed of fast and slow modes obtained for the system without transition region (compare Figure~\ref{fig:pctrdiag}(a) with Figure~\ref{fig:adiabatic}) since only a new isolated window of solutions ($\ctpc < \omega/k_x < \cspc$) is added to the phase speed diagram. This new window contains now a new kind of wave modes which we label PCTR slow modes, since they show a clear slow-like character and are dominated by  the PCTR physical conditions (see Section~\ref{sec:propctr}). An infinite number of kink and sausage harmonics is enclosed in this thin window. These new solutions are non-dispersive propagating waves in the transition region ($\kzpc^2 > 0$) but surface-like in the prominence ($\kzp^2 < 0$)

\begin{figure}[!h]
\centering
\includegraphics[width=0.85\columnwidth]{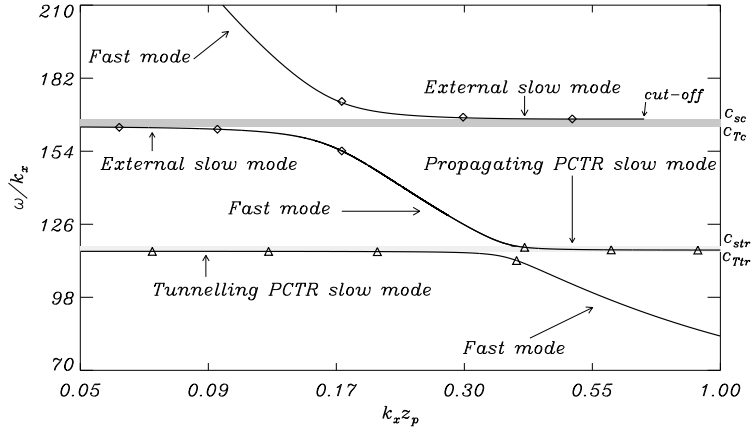}
\caption{Zoom of Figure~\ref{fig:pctrdiag}(b) to show the couplings of the fundamental fast kink mode (solid line) with the external slow kink mode (solid line and $\Diamond$) and the PCTR slow kink mode (solid line and $\triangle$). Note that the PCTR slow kink mode has a propagating branch ($\kzpc^2 > 0$) inside the region $\ctpc < \omega/k_x < \cspc$ (light-shaded region), and a tunnelling or evanescent branch ($\kzpc^2 < 0$). The dark-shaded region ($\ctc < \omega/k_x < \csco$) corresponds to the forbidden, leaky band. The phase speed is given in km~s$^{-1}$.} 
\label{fig:couplingpctr}
\end{figure}

\begin{figure}[!h]
\centering
\includegraphics[width=0.85\columnwidth]{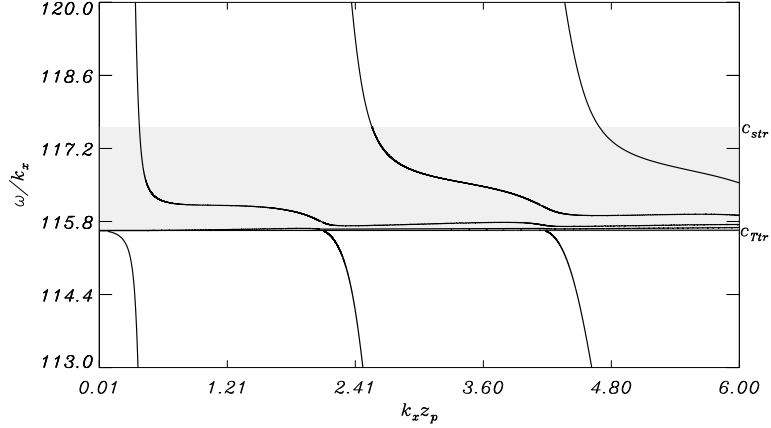}
\caption{Behaviour of the kink-like solutions inside the narrow region $\ctpc < \omega/k_x < \cspc$ (shaded zone) for a PCTR temperature of 500\,000~K. The quasi-horizontal lines correspond to PCTR slow modes whereas the dispersive lines are fast modes. Only several low harmonics are plotted, since the shaded region contains an infinite number of PCTR slow harmonics whose phase speeds cluster towards $\ctpc$. The phase speed is given in km~s$^{-1}$.} 
\label{fig:regionpctr}
\end{figure}

On the other hand, the situation $T_{\rm tr} > T_{\rm tr}^*$ has more physical consequences and, therefore, is investigated in more detail next (see Figure~\ref{fig:pctrdiag}(b)). For $T_{\rm tr} > T_{\rm tr}^*$ the PCTR slow modes contained in the region $\ctpc < \omega/k_x < \cspc$ interact with the fast modes by means of couplings. In this situation, the PCTR slow fundamental kink mode has an special relevance because its behaviour for small $k_x z_{\rm p}$ is different from that of the other PCTR slow modes. To illustrate this especial fact, a zoom to the dispersion diagram is performed in Figure~\ref{fig:couplingpctr}, displaying only the fundamental kink modes. There, we see that the fast mode and the PCTR slow mode show a coupling for $k_x z_{\rm p} \approx 0.4$. At the coupling, the curves show an avoided crossing and the oscillatory modes swap their magnetoacoustic properties. On the left-hand side of the coupling, the PCTR slow mode is a tunnelling solution in the PCTR ($\kzpc^2 < 0$), whereas on the right-hand side this solution is body-like ($\kzpc^2 > 0$). This peculiar behaviour of the PCTR slow fundamental kink mode is not observed for the other PCTR slow modes, which are always propagating waves enclosed in the region $\ctpc < \omega/k_x < \cspc$.  The behaviour of the PCTR slow harmonics can be seen in more detail in Figure~\ref{fig:regionpctr}. The phase speed of these PCTR slow harmonics clusters towards $\ctpc$, while the fast modes are dispersive and couple with the PCTR slow harmonics when their frequencies coincide. With respect to the prominence, all these solutions are body-like ($\kzp^2 > 0$).

In this way, we notice some parallelism in the behaviour of the PCTR slow fundamental mode and the external slow mode, since their phase speeds tend to cluster towards the tube speed of their respective dominant medium ($\ctc$ for the external mode and $\ctpc$ for the PCTR mode). Nevertheless, there is an important difference between their behaviours because the external slow mode cannot penetrate as a non-leaky solution in the region $\ctc < \omega/k_x < \csco$, and so its existence is restricted to small values of $k_x z_{\rm p}$. 

\subsection{Magnetoacoustic Properties of the New Oscillatory Modes}
\label{sec:propctr}

\begin{figure}[!h]
\centering
\includegraphics[width=\columnwidth]{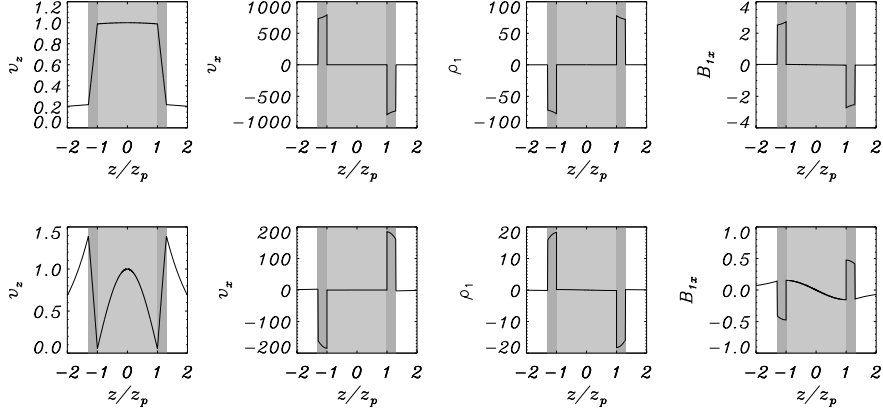}
\caption{Perturbations $v_z$, $v_x$, $\rho_1$, and $B_{1x}$ (in arbitrary units) versus the dimensionless distance to the prominence axis in the case of the PCTR slow fundamental kink mode. The top panels correspond to $k_x z_{\rm p} = 0.25$ (tunnelling branch) and the bottom panels to $k_x z_{\rm p} = 1$ (propagating branch). The PCTR temperature is $T_{\rm tr} =$~500\,000~K. The light-shaded region corresponds to the prominence slab and the dark-shaded zone to the PCTR.} 
\label{fig:eigenpctr}
\end{figure}

As we have just described, the presence of an isothermal PCTR produces a modification of the phase speed diagram and new solutions, the PCTR slow modes, are present.  To check their magnetoacoustic character, we fix the PCTR temperature to 500\,000~K (case $T_{\rm tr} > T_{\rm tr}^*$) and plot the eigenfunctions of the PCTR slow fundamental kink mode in Figure~\ref{fig:eigenpctr} for $k_x z_{\rm p} = 0.25$ (tunnelling branch) and for $k_x z_{\rm p} = 1$ (propagating branch). Perturbations are essentially confined within the PCTR in both cases and their amplitude is much larger in the PCTR than in the other regions, hence their behaviour is dominated by the PCTR physical conditions. The velocity perturbations present a well-defined polarisation, with the amplitude of $v_x$ much larger than that of $v_z$. In addition, the perturbations to the density are several orders of magnitude larger than those obtained for fast modes (see second and third rows of Figure~\ref{fig:autofunctions}). These facts are unequivocal evidence of the slow-like character of these modes. In the case $T_{\rm tr} < T_{\rm tr}^*$ the eigenfunctions are very similar to those displayed in Figure~\ref{fig:eigenpctr} and have not been plotted here for simplicity.

As we did for the internal slow modes, one can consider the approximation $v_{\rm ph} \approx \cspc$ to the phase speed of the PCTR slow modes. In this way, the dependence of the phase speed with the PCTR temperature is given by $v_{\rm ph} \sim T_{\rm tr}^{1/2}$. If a similar analysis to that of Section~\ref{sec:param} is performed, one can see that these solutions are very weakly affected by the value of the magnetic field strength and the coronal temperature.

%
%
%
\section{Discussion and Conclusions}

The investigation performed in the present paper is a simple study about the influence of the presence of a prominence-corona transition region on the magnetoacoustic oscillatory modes supported by the prominence body. The reader must be aware that the wave modes investigated here are not all the possible solutions supported by our equilibrium due to the restriction $k_y = 0$. For this reason, the case with $k_y \neq 0$ should be investigated in a future work. 

This study is divided into two parts. In the first one (Section~\ref{res:nopctr}), the JR92 prominence model for $k_y = 0$ (ER82 case) has been revised and the oscillatory modes have been reclassified according to their magnetoacoustic properties. This has allowed us to find a wave mode with slow-like properties (the external slow mode), which passed unnoticed in ER82 and JR92. In the second part of the paper (Section~\ref{res:pctr}), an isothermal PCTR has been considered in the system, and its effect on the previously existing wave modes has been assessed. In addition, a new solution has been obtained (the PCTR slow mode), which possesses slow-like properties and produces large longitudinal motions inside the transition region.

The main conclusions of this paper are summarised next.
\begin{itemize}

\item Internal slow modes are dominated by prominence conditions whereas the behaviour of the external slow mode is controlled by coronal physical conditions. The latter only exists as an evanescent-like solution for small wavenumber.

\item Fast modes are dominated by the prominence but show a slight dependence on coronal properties. The influence of the corona on fast modes grows for small $k_x$ (large wavelength) since perturbations are poorly confined.

\item The presence of an isothermal PCTR does not greatly affect the frequency of the oscillatory modes of the equilibrium without transition region if the PCTR sound speed is less than the internal Alfv\'en speed. In this situation, a new type of slow-like oscillatory modes (PCTR slow modes) which produce large displacements of the plasma inside the PCTR appear in an isolated window in the phase speed diagram. These new solutions do not interact with the previously existing solutions of the system without PCTR.

\item When an isothermal transition region with a tube speed greater than the internal Alfv\'en speed is considered, the PCTR slow modes interact with the previously existing fast modes by means of couplings.

\end{itemize}

The large longitudinal displacements of the plasma inside the transition region produced by the PCTR slow modes could be detected as oscillations in the intensity or Doppler shift of spectral lines associated with typical PCTR temperatures ({\it e.g.} \opencite{pouget}). As mentioned in the Section~\ref{Introduction}, there are very few observations of oscillations linked to prominences in transition region spectral lines, so one must be cautious about the observational values and their theoretical interpretation. Despite these remarks, the addition of the PCTR in theoretical investigations of prominence oscillations is important because it gives a way to understand observations obtained with wavelengths corresponding to different plasma temperatures. The consideration of an isothermal, homogeneous PCTR is a very rough approximation to the real situation but this simple theoretical study has allowed us to obtain some interesting results. From our point of view, the most important of all these results is that global fast modes of the system can interact with PCTR slow modes, which are essentially confined in the narrow PCTR. This fact could become of especial relevance if a more realistic, smooth PCTR is assumed. Then, a slow continuum would be present instead of individual PCTR slow modes. The interaction of the fast mode with this slow continuum could have important repercussions on the wave behaviour since a resonance phenomenon could occur. The consideration of a smooth PCTR is therefore the next step to be undertaken in future investigations, the present isothermal case being a simple first approach to the problem.


\acknowledgements

The authors acknowledge the financial support received from the Spanish Minis\-terio de Ciencia y Tecnolog\'ia and the Conselleria d'Economia, Hisenda i Innovaci\'o under grants AYA2006-07637 and PCTIB-2005GC3-03, respectively. R.~Soler thanks the Conselleria d'Economia, Hisenda i Innovaci\'o for a fellowship. The authors also thank the referee for their helpful corrections and comments.


\end{article} 
\end{document}